\documentclass[sigconf]{aamas}

\usepackage{balance} % for balancing columns on the final page
\usepackage[utf8]{inputenc} % allow utf-8 input
\usepackage[T1]{fontenc}    % use 8-bit T1 fonts
\usepackage{hyperref}       % hyperlinks
\usepackage{url}            % simple URL typesetting
\usepackage{balance}
\usepackage{booktabs}       % professional-quality tables
\usepackage{amsfonts}       % blackboard math symbols
\usepackage{nicefrac}       % compact symbols for 1/2, etc.
\usepackage{microtype}      % microtypography
\usepackage{xcolor}         % colors

\usepackage{graphicx}
\usepackage{amsmath}
\usepackage{amsthm}
\usepackage[ruled,vlined]{algorithm2e}
\usepackage{cancel}
\usepackage[T1]{fontenc}
\usepackage{newtxtext,newtxmath}

\DeclareMathOperator\supp{supp}
\newtheorem{theorem}{Theorem}

\newtheorem{proposition}{Proposition}

\newcommand{\ie}{{\it i.e.}, }
\makeatletter
\gdef\@copyrightpermission{
  \begin{minipage}{0.2\columnwidth}
   \href{https://creativecommons.org/licenses/by/4.0/}{\includegraphics[width=0.90\textwidth]{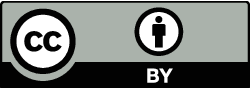}}
  \end{minipage}\hfill
  \begin{minipage}{0.8\columnwidth}
   \href{https://creativecommons.org/licenses/by/4.0/}{This work is licensed under a Creative Commons Attribution International 4.0 License.}
  \end{minipage}
  \vspace{5pt}
}
\makeatother

\setcopyright{ifaamas}
\acmConference[AAMAS '25]{Proc.\@ of the 24th International Conference
on Autonomous Agents and Multiagent Systems (AAMAS 2025)}{May 19 -- 23, 2025}
{Detroit, Michigan, USA}{Y.~Vorobeychik, S.~Das, A.~Nowé  (eds.)}
\copyrightyear{2025}
\acmYear{2025}
\acmDOI{}
\acmPrice{}
\acmISBN{}

\acmSubmissionID{fp0089}

\title[Learning in Games with Progressive Hiding]{Learning in Games with Progressive Hiding}

\author{Benjamin Heymann}
\orcid{0000-0002-0318-5333}
\affiliation{
  \institution{Criteo AI Lab}
  \city{Paris}
  \country{France}}
\email{b.heymann@criteo.com}

\author{Marc Lanctot}
\orcid{0000-0003-4159-2519}
\affiliation{
  \institution{Google DeepMind}
  \city{Montreal}
  \country{Canada}}
\email{lanctot@google.com}

\begin{abstract}
When learning to play an imperfect information game, it is often easier to first start with the basic mechanics of the game rules.
For example, one can play several example rounds with private cards revealed to all players to better understand the basic actions and their effects. Building on this intuition, this paper introduces {\it progressive hiding}, an algorithm that balances  learning the basic mechanics of an imperfect information game and satisfying the information constraints. 
Progressive hiding is inspired by methods from stochastic multistage optimization, such as scenario decomposition and progressive hedging.
We prove that it enables the adaptation of counterfactual regret minimization to games where perfect recall is not satisfied. Numerical experiments illustrate that progressive hiding  produces notable improvements in several settings.
\end{abstract}


\newcommand{\BibTeX}{\rm B\kern-.05em{\sc i\kern-.025em b}\kern-.08em\TeX}

\begin{document}

%%% The following commands remove the headers in your paper. For final 
%%% papers, these will be inserted during the pagination process.

\pagestyle{fancy}
\fancyhead{}

%%% The next command prints the information defined in the preamble.

\maketitle 

%%%%%%%%%%%%%%%%%%%%%%%%%%%%%%%%%%%%%%%%%%%%%%%%%%%%%%%%%%%%%%%%%%%%%%%%

\section{Introduction}

This paper shows how the learning process in games with imperfect information can be improved by  relaxing the information constraints with penalty-like methods.
For a game theorist, {\it games} are mathematical models to describe strategic interactions between  agents  (the players of the game).
Games, in their standard acceptance, which usually supports this definition, are often referred to as the ``Drosophila of Artificial Intelligence''~\cite{mccarthy90}.\footnote{The fruit fly is widely used in experiments due to its ease of cultivation in large numbers outside its natural habitat and its rapid reproduction cycle.}
This comparison with the fruit fly comes from the fact that games provide Artificial Intelligence (AI) researchers with standardized, easily reproducible environments where they can test and benchmark various AI techniques. Games  offer challenges that are complex enough to be interesting for AI research, but still have well-defined rules.
One of the longstanding AI research questions about games is how to learn good strategies. 

The field has advanced significantly due to breakthroughs where AI algorithms have outperformed human players in strategic games, illustrating the potent applications of AI in complex decision-making environments.
Notable examples include Chess, Go, Poker and  Stratego~\cite{silver2018general,schrittwieser2020mastering,moravvcik2017deepstack,brown2018superhuman,brown2019superhuman,bowling2015heads,perolat2022mastering}.

While Chess and Go are perfect information games, Stratego and Poker  are imperfect information games, in the sense that the state of the game is not fully observed by the players.
Hence players must account for the private information held by other players, as well as how their actions reveal their private information.
Notably, this makes room for elements like bluffing~\cite{southey2012bayes}.
Regret minimization at scale played a key role in recent achievements in AI for imperfect information games~\cite{moravvcik2017deepstack,brown2018superhuman,brown2019superhuman}.

The AI community recently identified the card game Hanabi as a challenge for AI~\cite{BARD2020103216}. 
What differentiates Hanabi from the previously mentioned games is that it is a cooperative game\footnote{the term same payoff games is also very often used}: all players have the same objective. The difficulty comes from the uncertainty of the draws and from the fact that the players cannot communicate freely. 
The problem of pursuing a common objective with limited communication is not new, and spans several fields, in particular economic theory, control, game theory, and machine learning~\cite{witsenhausen1968counterexample,doi:10.1137/0309013,li2009learning,lanctot2012no,nayyar2013decentralized,carpentier2015stochastic,hu2019simplified,Sokota_Lockhart_Timbers_Davoodi_D’Orazio_Burch_Schmid_Bowling_Lanctot_2021,heymann2022kuhn}.
Witsenhausen showed, with his landmark counterexample~\cite{witsenhausen1968counterexample}, how imposing communication constraints on a problem drastically changes the nature of the problem.

We present a general framework for dealing with information in games. 
The framework is best described with the following real life observation: when teaching a child how to play a game ---  like Poker or Hanabi --- with hidden information, one often starts with a few plays where all the cards are revealed, and then, as the child is learning, one starts hiding more and more information to make the game more challenging. 
This metaphor summarizes the idea of the main algorithm presented  in this paper. Because of this metaphor--- and also as a tribute to Rockafellar and Wets' {\it progressive hedging} algorithm~\cite{rockafellar1991scenarios}\footnote{See Section~\ref{sec:related} for related work.}--- we name the algorithm {\it progressive hiding}. 

A common assumption in games is that players remember what they see and what they do in the precise order of occurrence.
In his seminal work~\cite{kuhn1953extensive} on the tree representation of extensive form games, this assumption is what Kuhn referred to as \textbf{perfect recall}. 
One difficulty with a game such as Hanabi is that, if one wants to see the whole group of players as one player, a \textit{team}, then this  \textit{team} does not satisfy the perfect recall assumption because this one (big) player that constitutes the team is not allowed to remember the private information  from its past. This absence of perfect recall prevents us from using key learning techniques~\cite{lanctot2012no,NIPS2009_00411460} such as counterfactual regret minimization (CFR)~\cite{zinkevich2007regret}.
As we show, progressive hiding's modification of the game allows recovering some notion of perfect recall, which makes the new problem amenable to  CFR.
In an effort to bridge ideas from decentralized control theory and computational game theory, we depart from the customary tree notations and introduce a hybrid between product games \cite{heymann2022kuhn} and games on trees. This new, compact notation reveals quite practical for our purpose. While product games were fully formalized relatively recently, they are just the continuation of an idea that was hinted half a century  ago by Witsenhausen~\cite{doi:10.1137/0309013}, specifically as a way to deal with setting such as ours (communication constraints).

\subsection{Contribution}
Our main contribution -- Theorem~\ref{th:CFR} -- is to show how information relaxation can be made compatible with learning in games. 
Information relaxation has never been used in imperfect information games, despite the fact that it is very close to how people discover how to play games in real life. 

While we mostly focus on  same payoff settings in this paper, we believe progressive hiding and information relaxation methods to be full of potential for the competitive case too (as opposed to same payoff games). The reason is similar to the one provided in the paper on progressive hedging~\cite{rockafellar1991scenarios} for solving stochastic dynamic programs: take a game for which a large part of the complexity comes from the hidden information (think Stratego or Poker or Bridge). Without the hidden information, those games could be amenable to deterministic methods (dynamic programming, for instance). Progressive hiding provides a way to navigate between the two worlds (deterministic and stochastic). In this work, we deliberately focus on same payoff games because we believe Theorem~\ref{th:CFR} illustrates well the strength of the general philosophy of information relaxation.

Section~\ref{sec:preliminary} introduces some background, including the product game notation, that will prove practical for our purpose. Section~\ref{sec:related} is an attempt to make a historical connection between algorithmic game theory and a class of stochastic programming technique called scenario decomposition (e.g. progressive hedging). 
Section~\ref{sec:information-relaxation} presents a penalty method for games that can be analyzed under the lens of the proximal algorithms. This method is a first step toward our proposal, and brings the optimization tools that will be assembled with the learning ones in the second step. 
We then introduce in Section~\ref{sec:progressive-hiding} progressive hiding (Algorithm~\ref{algo:progressive-hiding}), an algorithm that combines no-regret learning and information-relaxation. 
We show, in particular, how this approach allows us to recover the main property of CFR in an auxiliary game. 
We document numerical experiments  in Section~\ref{sec:experiments}. Section~\ref{sec:conclusion} provides discussion on the  limits of Algorithm~\ref{algo:progressive-hiding}. 
All proofs and code are provided in the \href{https://www.arxiv.org/pdf/2409.03875}{arXiv} version of the paper.

\section{Preliminary}
\label{sec:preliminary}

This section is dedicated to background material on learning in games.
There are different ways of modeling extensive form games~\cite{ritzberger2016theory}, typically modeled as game trees. 
In this paper, we borrow ideas from the product form~\cite{heymann2022kuhn} as it better fits our needs for modeling the information structure.  Games in product form represent information  over a product set.

An important property of product games is that modifying the information structure of a product game can be done without modifying the product set over which this product game is defined. Such property is not satisfied by the tree representation (see Appendix on \textit{lumberjacking}).
Also, the product form provides a common framework for game theory and causality theory~\cite{heymanncausal}. 

In what follows, if $S$ is a set and $s\in S$, we will use the notation $-s$ to refer to the set $\{s'\in S,s'\neq s\}$ when it is clear what $S$ is from the context.
If $n$ is an integer, then we set $[n] = \{1, 2, \ldots, n \}$.
\subsection{Motivation for using the  product form}

In an extensive form game (EFG), the order of play is encoded in a tree structure. 
This notation originates from~\cite{kuhn1953extensive} and is used extensively in game theory. 
In this paper, we use the more general formalism of product games.
 Product games shift the focus from the temporal order of play to the information available at each decision nodes. The game is supported by a Cartesian product instead of being supported by a tree. The reason to use this generalization of EFG is the need for flexibility when we modify the information structure of the game when doing information relaxation.  

For example, suppose we discover at the end of the game if a coin toss was head or tail. Then we might end-up with a  different tree representation when we allow the player to observe the coin toss result at the beginning of the game.
Similarly, if one consider a game with hidden actions, and several possible variations where some of those hidden actions are revealed. Then the resulting trees will differ in their structures. For instance, if Bob and Alice play Rock-Paper-Scissors, a tree that represent a modified game where Bob observes Alice action (Alice plays first,   Bob second) will not be the same as a tree where Alice sees Bob action (Bob plays first,   Alice second).

In this work, the product formalism allows to abstract away from the order of play encoded in the tree representation. It also allows defining useful objects such as information maps and projectors in a very compact manner.
Compared to general extensive-form games as in ~\citet[Chapter~11]{Osborne1994}, Nature plays \textit{only once} at the very beginning, and the random state is revealed progressively to the players. The order of play does not need to be known in advance, as it is the case for games on tree. 
This choice of formalism is made without loss of generality and based on convenience for the problem considered in this paper.

\subsection{Extensive Game in Product Form}
\label{seq:extensivegameinproductform}
We first introduce a minimalistic version of the product form \footnote{without explicit references to the \textit{information algebras} and the \textit{solution map}} for extensive form games, which is an alternative to the tree representation.
The appendix contains more comprehensive information on the product game formalism.  
In what follows, $P$ is  the set of players in the game, $L$ is a positive integer that represents the maximum number of turns in the game (\ie maximum length of any game  ), and $W$ is a positive integer that represents the maximum number of legal actions at any time. 
We denote by $\Delta$ the $W$-dimensional simplex.
We encode the available information to a player  with an abstract, finite set 
 $\mathbb{G}$. Otherwise said, $\mathbb{G}$  is  a finite set that is used to encode the players' information as the game proceeds. 
The randomness owned by the game itself (as opposed to the randomness controlled by the players), sometimes referred to as \textbf{Nature's moves}, is encoded with 
 a discrete probability space $(\Omega,\mathbb{P})$.
 The set \(\Omega\) represents the space where random events take place. For example, in a card game, \(\Omega\) would be the set of all possible orders of cards in the deck. In a dice game, it would be the set of all possible sequences of dice rolls during the game. There is a probability measure on \(\Omega\), denoted by \(\mathbb{P}\). For instance, in Poker, all possible orders of cards are equally probable.
 Last we define the \textbf{product set}  as $\mathbb{H}=\Omega\times[W]^{L} $.

Given the primitives already introduced, a  \textbf{game in product form }
consists of a tuple $(\mathcal{P},\mathcal{A},\mathcal{X},r)$ where: 
$\mathcal{P}:[L]\to P$ is a map that indicates which player needs to play at stage $i\in [L]$; $\mathcal{A}:\mathbb{G}\to [W]$ is a map that encodes the number of available actions at each stage of the game; for all $i\in [L]$, a map $\mathcal{X}_i :\mathbb{H} \to\mathbb{G}$ that encodes the available information at each stage of the game, and that only depends  on the first $i$ components of $\mathbb{H}$ (Nature's move and the predecessors);\footnote{this assumption can be weakened, but helps the presentation}  the map $r:\mathbb{H}\to \mathbb{R}^P$ encodes the reward of the game, more precisely, $r_p$ is the reward of player $p$.
For any element in $\mathbb{H}$, we denote by $h_{\emptyset}$ its first component (associated with Nature's moves) and by $h_i$ the $(i+1)^{th}$ component, associated with the $i^{th}$ move of the players.

\subsection{Example of game in product form}

The product form for Cooperative Matching Pennies (Figure~\ref{fig:matching_pennies_tree}) involves one (team)-player represented by Alice and Bob.
Nature has two equiprobable moves in $\Omega=(=, \neq)$.
Alice and Bob pick their decisions in $(H, T)$ and $(H, T, P)$, respectively.
So the product of Nature's set and the action sets is $\mathbb{H}=(=, \neq) \times(H, T) \times(H, T, P)$.
Bob only observes  Alice's move. Formally, for

\begin{align*}
  h_1=\left(\omega^1, d_{\text {Alice }}^1, d_{\text {Bob }}^1\right) \quad \text { and } \quad h_2=\left(\omega^2, d_{\text {Alice }}^2, d_{\text {Bob }}^2\right),  
\end{align*}

we have
$\mathcal{X}_{\text {Bob }}\left(h_1\right)=\mathcal{X}_{\text {Bob }}\left(h_2\right)$ if $d_{\text {Alice }}^1=d_{\text {Alice }}^2$, and since Alice only sees Nature's decision:
$\mathcal{X}_{\text {Alice }}\left(h_1\right)=\mathcal{X}_{\text {Alice }}\left(h_2\right) \Longleftrightarrow \omega^1=\omega^2$. We do not care about the codomain of $\mathcal{X}$ as long as we can encode this type of relation. The seminal paper on product games~\cite{heymann2022kuhn} does not use the information map $\mathcal{X}$, but this requires manipulating more abstract objects (e.g., $\sigma$-fields).
Alice's policy on $\mathbb{H}$ can be reduced to a function on $\Omega$ via the information map.
Similarly, Bob's policy can be represented by a function whose domain is Alice's decision set.
Given such policies, sampling from $(=, \neq)$, Alice's policy, and Bob's policy produces a sequence in $\mathbb{H}$ with distribution $Q_\mu$ that we present in Section~\ref{seq:pushforward}.

\subsection{Policies and Push-Forward Probability}
\label{seq:pushforward}
Next, building on the primitives we  introduced in Section~\ref{seq:extensivegameinproductform}, we can now describe how a product game is played by specifying the policies and the resulting push-forward probability.
For a decision time $i\in [L]$, we denote by $\bar{\Lambda}_i$ the set of  (admissible) policies $\mu_i$, where 
$\mu_i:\Omega\times[W]^{i-1}\to \Delta $ and $
    \supp{\mu_i(h)}\subset [\mathcal{A}\circ \mathcal{X}_i(h)]$,
that is, the policy only selects admissible actions.
Using a canonical extension, we can identify $\mu_i$ with a map from $\mathbb{H}\to \Delta$ that  depends at most on Nature's move and the first $i-1$ decisions. 
For $i\in [L]$, an \textbf{implementable}  policy  $\mu_i\in \bar{\Lambda}_i$ should also satisfy, for any histories $h$ and $h'$ in  the product space $\mathbb{H}$
\begin{align}
\mathcal{X}_i(h)=\mathcal{X}_i(h')\implies\mu_i(h)=\mu_i(h'). \label{eq:policy1} 
\end{align}
In words, an implementable policy depends solely on what player $\mathcal{P}(i)$ is supposed to know at the time of the decision. 
Relation~\eqref{eq:policy1} is the \textbf{non-anticipativity constraint}~\cite{shapiro2021lectures}, and  indicates that the policy $\mu_i$ should only depend on the information given by $\mathcal{X}_i$ to player $\mathcal{P}(i)$ at step $i$.
If a policy is not implementable, it means that it needs some pieces of information that are not available at the time of the decision. 
We denote by $\Lambda_i$ the subset of policies from $\bar{\Lambda}_i$ that are implementable.
Also, we denote by $\Lambda_p$ and $\bar{\Lambda}_p$ the sets of implementable and  admissible policy profiles of player $p$ for $p\in [P]$: $\Lambda_p=\times_{i,\mathcal{P}(i)=p}\Lambda_i$. We similarly define $\Lambda$ and $\bar{\Lambda}$ to refer to policy profiles of all the players: $\Lambda=\times_{p}\Lambda_p$.

The detail of the construction of a game in product form goes back to Witsenhausen's seminal paper, and was then clarified in ~\citet{doi:10.1137/0309013,heymann2022kuhn}. We recognize that the construction might feel uneasy for readers used to the tree formulation.
We refer to~\cite{doi:10.1137/0309013,heymann2022kuhn} for more details on this  construction.

How do we play a game in product form? Given a deterministic policy profile $\mu$ and an element $\omega$ of $\Omega$, there is a unique   element $h$ of $\mathbb{H}$ that satisfies $(\omega,\mu(h))=h$.    We pinpoint that this relation  is on the full strategy profile (on all the players, not only on a player of interest). The relation states that if we look at a realization of a game (all the decisions), and replay the decisions along the path using the strategies that were used for this realization, we should recover the realization itself. It is a well-posedness condition. Typically, one wants to exclude temporal paradoxes.
In a game on tree, one usually first describes the game with deterministic policies and then specifies a way for the players to randomize. Here it is the same, the well-posedness of the game is checked on a deterministic specification. Then~\cite{heymann2022kuhn}  provide constructions to extend the space of pure strategies to randomized strategies.

Given a policy profile
$(\mu_i)_{i\in [L]}$, we can therefore construct the associated push-forward probability\footnote{The term push-forward comes from probability theory. }, denoted by $\mathbb{Q}_{\mu}$, on the product space $\mathbb{H}$. 
It is a probability on the realizations of the game. It obviously depends on the strategies of the players $\mu$.
We then denote by $\mathbb{E}_{\mu}$ the \textbf{associated expectation operator}.
Last, we define a notation for modifying policies that allows for easy adjustment of specific components. Let \(\mu\) be a base policy, and consider the operation of altering its \(i\)-th component to a new value \(k\). This modification is denoted by \(\mu{(i \to k)}\). \footnote{Formally, the modified policy \(\mu(i \to k)\) is given by
\[
\mu(i \to k)_j = 
\begin{cases} 
k & \text{if } j = i, \\
\mu_j & \text{otherwise}.
\end{cases}
\]
This notation can be extended to accommodate multiple simultaneous modifications. For instance, changing the \(i\)-th component to \(k\) and the \(j\)-th component to \(l\) in \(\mu\) is expressed as \(\mu(i \to k)( j \to l)=\mu(i \to k, j \to l)\). }

An $\epsilon$-Nash equilibrium is an admissible strategy profile $\mu$ so that no player can improve her outcome by more than $\epsilon$ from an unilateral deviation, that is, for any player $p$ and any other admissible strategy profile $\mu'_p$, 
$
    \mathbb{E}_{\mu_p,\mu_{-p}}[r_p(\textbf{h})]+\epsilon\geq \mathbb{E}_{\mu'_p,\mu_{-p}}[r_p(\textbf{h})].
$
A Nash equilibrium is a $0$-Nash equilibrium. 

\subsection{Perfect Recall and Information Maps}
\label{sec:information-map}
Perfect recall is a standard assumption in extensive-form games~\cite{ShohamLB09,zinkevich2007regret}. An important property of perfect recall is that it implies the equivalence between mixed and behavioral strategies~\cite{kuhn1953extensive,heymann2022kuhn}.
In words, perfect recall  means that the player has perfect memory of what  they see and do and the precise order that information was revealed over the turns. 
In our context, perfect recall for player $p$ corresponds to the two conditions, for any  $i<j$ such that $\mathcal{P}(i)=\mathcal{P}(j)=p$
\begin{align}
\label{eq:perfect-recall}
    \bigg(\mathcal{X}_{i}(h)\neq\mathcal{X}_{i}(h')&\implies
    \mathcal{X}_{j}(h)\neq\mathcal{X}_{j}(h') \bigg)\\ \text{ and }   \bigg(h_i\neq h_i'&\implies
    \mathcal{X}_{j}(h)\neq\mathcal{X}_{j}(h') \bigg).
\end{align}
The first condition states that if the information known about $h$ and $h'$ differs on turn $i$, then this must remain true at a future turn $j > i$.
The second condition states that if the action taken at turn $i$ is different, then the player must remember this information at a later turn $j > i$.
While the perfect recall assumption is realistic when the player corresponds to a single agent, this is less so when the player is used to model a team of several agents.  The absence of perfect recall in a game prevents the use of several tools, for instance,  backward induction, CFR~\cite{zinkevich2007regret,lanctot2012no}, and the equivalence of behavioral and mixed policies~\cite{kuhn1953extensive}.

We say that an information map $\mathcal{X}_i^+$ is \textit{finer} than an information map $\mathcal{X}_i$ if for any $h\in \mathbb{H}$
       $ \mathcal{X}_{i}(h)\neq\mathcal{X}_{i}(h')\implies
    \mathcal{X}_{i}^+(h)\neq\mathcal{X}_{i}^+(h') $.
    If $\mathcal{X}_i^+$ is finer than $\mathcal{X}_i$, we can also say equivalently that 
    $\mathcal{X}_i$ is \textit{coarser} than $\mathcal{X}_i^+$.
We say that $\Lambda^+\subset\bar{\Lambda}$ \textit{is induced} by an information map profile $\mathcal{X}^+$ if it is the set of policy profiles from $\bar{\Lambda}$ that are implementable with respect to $\mathcal{X}^+$.
When we take a finer information map, the game become in some sense easier, because there is less uncertainty. 
The idea  of information relaxation is to take a finer information map to help the learning process, and then penalize the usage of the free information provided. An important tool to do so is the projector, which we define next.

\subsection{Projector}

\label{sec:proj}
Let $\mu$ be any full support implementable policy profile, and $\tilde{\mathbb{H}}$ be the support of $\mathbb{Q}_\mu$.
For $\mu_0\in\bar{\Lambda}$ of full support,   we define  $\mathrm{Proj}_{\mu_0}(\mu)$ as the element $\gamma$ of $\bar{\Lambda}$ such that 
\begin{align}
\label{eq:projection}
    \gamma_i(h) =  \mathbb{E}_{\mu_0}[
    \mu_i(\textbf{h'})|\mathcal{X}_i(\textbf{h'}) = \mathcal{X}_i(h)]
    \quad \forall i\in [L]\quad \forall h\in\mathbb{\tilde{H}}.
\end{align}
Suppose $\mu_i$ was defined using an information map finer than $\mathcal{X}$, then, by property of the condition expectation,  the projector transforms $\mu_i$ into an implementable policy for $\mathcal{X}_i$. 

We have the following key properties (adapted from~\cite{rockafellar1991scenarios}):
\begin{proposition}
\label{prop:projection_admissibility}
    $\forall \mu\in\bar{\Lambda}, \mu\in\Lambda\iff \mathrm{Proj}_{\mu_0}(\mu)=\mu$.
\end{proposition}

\begin{proposition}
\label{prop:proj_indempotent}
   $\mathrm{Proj}_{\mu_0}\circ \mathrm{Proj}_{\mu_0}=\mathrm{Proj}_{\mu_0}$.
\end{proposition}

Notably, proposition~\ref{prop:projection_admissibility} provides an \textbf{alternative representation of the non-anticipativity} constraint~\eqref{eq:policy1}.

\subsection{No-regret learning in games}
\paragraph{No-regret learning}
The theory of online learning envision a decision maker that makes sequential decisions $x_t$ in a convex, compact set $\mathbb{X}$ at each epoch $t=1\ldots T$, afterwhat the (possibly adversarial) environment outputs a concave reward $\ell_t$. An important metric of success for an algorithm is the regret, defined as
$
 \max_{x\in\mathbb{X}}\sum_{t=1}^T  \ell_t (x)  - \sum_{t=1}^T \ell_t (x_t).
$
A regret minimization algorithm is one where this regret grows at a rate of $o(T)$.
An algorithm with this property is called\textit{ Hannan-consistent}.
Regret minimization algorithms  play a central role in optimization and game theory, because they can be used to maximize a function or find approximate Nash equilibrium~\cite{Blum07}. For example, in the zero-sum two-player setting, one can find $\epsilon$-Nash equilibrium by repeating the game with two Hannan-consistent learners.

\paragraph{CFR}
We next describe counterfactual regret minimization (CFR), which was introduced in~\cite{zinkevich2007regret} and extended in many papers, in particular~\cite{NIPS2009_00411460,johanson2012efficient,lisy2015online,lanctot2012no}.
For $k\in [W]$ denote by $\delta_k$ the constant policy that always select action $k$.
The CFR algorithm keeps a collection of local regret minimizers indexed by 
$\{(i,g)\in [L]\times \mathbb{G}; g\in \mathcal{X}_i(\mathbb{\tilde{H}})\}$. Then at each step, the regret minimizer $(i,g)$ is fed with the reward vector of 
 size $[\mathcal{A}(g)]$ and of kth component $\mathbb{E}_{\delta_k, \mu_{-i}^{t}}[r_{\mathcal{P}(i)}(\textbf{h})|\mathcal{X}_i(\textbf{h}) = g]$ for $k \in [\mathcal{A}(g)]$. The strength of CFR is that when a player satisfies the perfect recall assumption~\ref{eq:perfect-recall}, their overall regret is upper-bounded by the  sum of the regrets of the local regret minimizers.

\section{Related Works}
\label{sec:related}
In this section, we attempt to draw  a connection between algorithmic game theory and a stochastic programming technique called scenario decomposition.
As argued in~\cite{rockafellar1991scenarios} a common methodology for addressing uncertainty involves initially adopting a deterministic simplification. Practitioners first solve the problem for some known scenarios, thereby establishing a baseline of deterministic solutions. Subsequently, these solutions are aggregated to incorporate uncertainty. This approach simplifies complex problems by breaking them down into more manageable components.

In game theory, this  type of approach is sometimes branded as  Perfect Information Monte Carlo search (PIMC search~\cite{long2010understanding}).
Motivated by the unexpected success of the methods in Bridge~\cite{ginsberg2001gib}, Skat~\cite{buro2009improving} and Hearts~\cite{sturtevant2008analysis},
the authors of~\cite{long2010understanding}, discuss conditions for PIMC to work. While the theoretical limit of the PIMC approach has been recognized very early~\cite{frank1998search}, it still achieves state of the art performance in games such Skat and Bridge.
Further improvements of PIMC are proposed in~\cite{furtak2013recursive,solinas2019improving}.

Paraphrasing the authors of~\cite{rockafellar1991scenarios}, the aim of progressive hedging is to provide  "a rigorous algorithmic procedure for determining (...) a policy in response to any weighting of the scenarios". 
The  algorithm iteratively tweaks the rewards in each scenarios so that at some point, the procedure finds a policy that is adapted to the uncertainty filtration. 
The algorithm leverages the ideas of augmented Lagrangian and proximal algorithms, and enjoys theoretical guaranties under some convexity assumptions.  Progressive hedging was recently adapted to variational inequalities~\cite{rockafellar2019solving}. In the same vein, the recent monograph~\cite{shapiro2021lectures} presents Lagrangian methods applied to non-anticipativity constraints in the context of stochastic programming. The discretization of the random space and the
introduction of stochastic dual variables allows decomposing stochastic programs per scenarios. The dual interpretation comes with important results from convex analysis~\cite{rockafellar1997convex}.

Crucially,  with these stochastic programming techniques, the decomposition is done per scenario, not per agent. To apply online learning techniques, we would prefer to obtain a decomposition per agent. 
Furthermore, in  stochastic optimization,  a  sole decision maker optimizes jointly all the decisions. By contrast, in  games with imperfect information, 
players do not observe the same information, and 
the action of a player can influence what the other players observe.

\section{Information Relaxation}
\label{sec:information-relaxation}
In this section, we combine the penalty method and proximal algorithms~\cite{parikh2014proximal} to produce a player best response, that is, a policy that is optimal given the policy of the other players.  
While the resulting Algorithm~\ref{algo:RIR} might be of independent interest, it serves as a guiding principle for the design of progressive hiding, presented in the next section. 
We  first reduce to the case where $|P|=1$ by fixing the policies of all players but one.
Let $\tilde{\Lambda}\subset\bar{\Lambda}$ induced by an information map profile $\mathcal{\tilde{X}}$ such that $\Lambda\subset\tilde{\Lambda} $. Instead of restricting our search to implementable policies, we could  optimize over $\tilde{\Lambda}$. However, the optimal policy would very likely not satisfy the non-anticipativity constraint. 
Using Proposition~\ref{prop:projection_admissibility}, we suggest to favorize implementability through  a penalty term $-\lambda ||\mu-\mathrm{Proj}(\mu)||^2$, where $\mathrm{Proj}$ is a projector as introduced in Section~\ref{sec:proj} induced by a policy $\mu_0$ and $||\cdot||$ is the reweighted $L_2$ norm associated with $\mu_0$.
This idea drives us to a relaxed version of the problem of finding an implementable best response. Indeed, setting
$
       \mathcal{L}(\mu,\lambda)= \mathbb{E}_{\mu}[r_p(\textbf{h})]-\lambda||\mu-\mathrm{Proj}(\mu)||^2,
$
we propose to solve the relaxed problem
\begin{align}
\label{eq:criterion}
   \max_{\mu\in \tilde{\Lambda}} \mathcal{L}(\mu,\lambda).
\end{align}
Observe that, because of the penalty, the criterion  is unusual and  might be impractical for many game solving approaches. A first step to address this difficulty is Theorem~\ref{th:RIR}, which states that one can produce a local optimum by composing two proximal steps, which can be analyzed under the lens of the Majorize-Minimization Algorithm  (Algorithm~\ref{algo:RIR}).
Progressive hiding builds on this idea by adding a linearization step as in~\cite{zinkevich2003online} for online convex problems.

\begin{algorithm}
\caption{Resolution by Information Relaxation}
\label{algo:RIR}
\DontPrintSemicolon  % Suppress semicolon ending on each line
\SetAlgoLined  % Enables vertical lines indicating block level
\SetKwInput{KwInput}{Input}  % Set the Input keyword
\SetKwInput{KwOutput}{Output}  % Set the Output keyword
\SetKwInput{KwInitialization}{Initialization}  % Set the Initialization keyword
\SetKwFor{For}{for}{do}{endfor} % Define repeat structure
\SetKw{KwReturn}{return}  % Define 'return' keyword

\KwInput{$\lambda>0$, $\mu^{(0)}\in \bar{\Lambda}$, $T\in\mathbb{N}$}
\KwInitialization{$\mu = \mu^{(0)}$}
\For{$t \leftarrow 1$ \KwTo $T$}{
    $\gamma \leftarrow \text{Proj}(\mu)$\tcp*{projection step} \;
    $\mu \leftarrow \arg\max_{\mu\in \tilde{\Lambda}} \mathbb{E}_{\mu}[r_p(\textbf{h})] - \lambda||\mu - \gamma||^2$ \tcp*{proximal step}\;
}
\KwReturn{$(\mu, \gamma)$}
\end{algorithm}

\begin{theorem}
\label{th:RIR}
Fix $\lambda>0$, $\mu^{0}\in\bar{\Lambda}$ and denote by 
$(\mu^T,\gamma^T)$
the output of    Algorithm~\ref{algo:RIR}, then 
 $\mathcal{L}(\mu^T,\lambda)$ is non-decreasing and admits a limit.
\end{theorem}

\begin{proposition}
\label{prop:upper_bound} We have the relation
\begin{align}
    \max_{\mu\in \Lambda} \mathbb{E}_{\mu}[r_p(\textbf{h})]\leq   \max_{\mu\in \bar{\Lambda}} \mathbb{E}_{\mu}[r_p(\textbf{h})]-\lambda ||\mu-\mathrm{Proj} (\mu)||^2.
\end{align}
\end{proposition}
The type of upper bound of Proposition~\ref{prop:upper_bound} is often used to estimate duality gap in operations research~\cite{brown2022information}.
Observe that when $\lambda$ is small, Problem~\eqref{eq:criterion} becomes closer to the problem of finding an optimal policy in $\tilde{\Lambda}$, so the couple $(\mu,\gamma)$ produced by Algorithm~\ref{algo:RIR} might not satisfy $\mu$ being close to $\gamma$, which means that  $\mu$ might be far from implementable.
By contrast, if $\lambda$ is large, one expects $\mu$ and $\gamma$ to be close, however, because of the proximal term, in each iteration only solutions very close to the previous steps will be considered, and we might get stuck with a local optima. 
In practice, many strategies could be used to solve the $\arg \max$ step in Algorithm~\ref{algo:RIR}. In particular, if  the auxiliary game (with $\tilde{\Lambda}$ instead of $\Lambda$) satisfies perfect recall, then one can rely on backward induction.
The next section, which presents our main contribution, is guided by the idea of combining the philosophy of Algorithm~\ref{algo:RIR} with regret minimization approaches.

\section{Progressive Hiding}
\label{sec:progressive-hiding}

We next present 
an algorithm ---
progressive hiding --- that replicates the philosophy   of information relaxation in the framework   of no-regret learning by introducing an auxiliary game.
This section also contains the article principal contribution, which is Theorem~\ref{th:CFR}. 
We suppose the players in $-p=P\setminus \{p\}$ play according to a policy profile $\mu^t_{-p}$ at stage $t$.

\subsection{Algorithm}

We suppose we have access to a class of \textbf{low-regret  algorithms}    that implements two functions: \textsc{Observe} (to get the realization of vector of linear reward) and  \textsc{Decide} (to output a probability distribution on the decision). 
Progressive hiding requires at each time $t$ \textbf{a non-negative penalty parameter} $\lambda^{t}$ for  the non-anticipativity constraint violation. This sequence of parameters $\lambda^{t}$  might be time adaptive  to allow for more flexibility at the beginning, but still shift toward implementability at the end of the learning process. 
Progressive hiding also requires an \textbf{information map}
$\tilde{\mathcal{X}}$, which is a refinement (cf. Section~\ref{sec:information-map}) over the  information map $\mathcal{X}$. For example, $\tilde{\mathcal{X}}$ can inform the player about the hands of its opponents, or about what other teammates saw in the previous turns. The finer the information map $\tilde{\mathcal{X}}$, the more additional information is provided in the auxiliary game.  Notably, ensuring that perfect recall is satisfied in the auxiliary game seems to be  a fruitful direction, as indicated by Theorem~\ref{th:CFR}.
Last, progressive hiding also requires  a sequence of \textbf{projectors}  $\text{Proj}^t$ as defined in Section~\ref{sec:proj}.
We set $\mathrm{Proj}^t=\mathrm{Proj}_{\mu^t_p,\mu^t_{-p}}$.
We denote by $I_p$ the elements $i$ of $[L]$ such that $\mathcal{P}(i)=p$, 
and we set $G_i = \tilde{\mathcal{X}}_i(\mathbb{H})$, for $i\in I_p$. We introduce the notation $\mathbb{E}_{\mu}[\textbf{r}_p|g]=\mathbb{E}_{\mu}[r_p(\textbf{h})|g\in\{\tilde{\mathcal{X}}_i(\textbf{h}),i\in I_p\} ]$.
Last, 
we introduce the time-dependent, \textbf{random penalty}, for $i\in I_p$ and $h\in \tilde{\mathbb{H}}$
\begin{align}
\label{eq:random_penalty}
   \ell^t_i(\mu_i(h),\tilde{\mathcal{X}}_i(h))= \lambda^t||\mu_i(\tilde{\mathcal{X}}_i(h))-\text{Proj}^t(\mu^t_p)_i(\tilde{\mathcal{X}}_i(h))||^2.
\end{align}
The right-hand side of Equation~\eqref{eq:random_penalty}  makes sense because (1) $\mu_i$ is a function of $\tilde{\mathcal{X}}_i(h)$, (2) $\text{Proj}^t(\mu^t_p)$ is a function of $\mathcal{X}_i(h)$ and $\tilde{\mathcal{X}}$ is finer than $\mathcal{X}$. Equation~\ref{eq:random_penalty} corresponds to a penalization of the\textit{ lack of implementability }of the current solution.

Let $\mu^{t}$ be the iterates of progressive hiding defined in Algorithm~\ref{algo:progressive-hiding}.
We denote   respectively by 
\begin{align*}
 \varrho_t(\mu_p) &= \mathbb{E}_{\mu_p,\mu_{-p}^t}\bigg[r_p(\textbf{h})   -  \sum_{i\in I_p}
 \ell^t_i(\mu_i(\textbf{h}),\tilde{\mathcal{X}}_i(\textbf{h}))\bigg],\\
   R^T &=\nicefrac{1}{T} \max_{\mu\in \tilde{\Lambda}}\sum_{t=1}^T\varrho_t(\mu_p)-\varrho_t(\mu^{t}_p),
   \end{align*}
   the new criteria for the player of interest at a given time step $t$,  and
the average regret $R^T$ in the auxiliary game with relaxed information constraints and penalty for violation. Also, for $i\in I_p$ and $g\in G_i$, we set
\begin{align*}
    \varrho_t[i,g][\mu_i]  = \mathbb{E}_{\mu^t_p(i\to \mu_i),\mu_{-p}}\Bigg[r_p(\textbf{h}) \\- \lambda^t\bigg(2\langle\mu_i^t(\textbf{h})-\textsc{Proj}^t(\mu^t_p)_i(\textbf{h})\mid\mu_i \rangle\bigg)-\sum_{i'>_p i} \ell^t_{i'}(\mu_{i'}^t(\textbf{h}),\tilde{\mathcal{X}}_{i'}(\textbf{h}))\bigg| g\Bigg],
\end{align*}
which corresponds to the \textit{local} criterion optimized by progressive hiding,  and $R^T_{loc}(i,g)=\nicefrac{1}{T} \max_{\mu_i\in \tilde{\Lambda}_i}\sum_{t=1}^T\varrho_t[i,g](\mu_i)-\varrho_t[i,g](\mu^{t}_i)$
 the local regret associated with this local criterion.
Remember that the projection can  also be regarded as a weighted average or a conditional expectation. The last term corresponds to the future quadratic penalties the player will incur on the current scenario if policy $\mu_i$ is chosen and the other policy components are kept fixed.
We also set $ R^{T,+}_{loc}(i,g)=  \max\left(R^{T}_{loc}(i,g),0\right)$.

Progressive hiding is displayed in Algorithm~\ref{algo:progressive-hiding}.
\begin{algorithm}
\caption{Progressive Hiding}
\label{algo:progressive-hiding}
\DontPrintSemicolon  % Suppress semicolon ending on each line
\SetAlgoLined  % Enables vertical lines indicating block level
\SetKwInput{KwInput}{Input}  % Set the Input keyword
\SetKwInput{KwOutput}{Output}  % Set the Output keyword
\SetKwFor{For}{for}{do}{endfor} % Define for loop
\SetKw{KwTo}{to}  % Define 'to' keyword for For loops
\SetKwInput{KwInitialization}{Initialization} 
\SetKw{KwReturn}{return}
\KwInitialization{Initialize  $\text{RegrMin}[i,g]$ for $(i, g) \in I_p \times G_i$}
\For{$t \leftarrow 1$ \KwTo $T$}{
    \ForEach{$(i, g) \in I_p \times G_i$}{
        $\mu^{t}_i(g) \leftarrow \text{RegrMin}[i,g].\textsc{Decide}()$ \tcp*{local no-regret  decisions}\;
    }
    $\gamma \leftarrow \textsc{Proj}^t(\mu^{t}_p)$  \tcp*{Projection step}\;
    \ForEach{$(i, g) \in I_p \times G_i$}{
    $\theta_{i,g}\leftarrow(\varrho_t[i,g] [\delta_d])_{d\in[\mathcal{A}_i(g)]}$\tcp*{reward in the auxiliary game}\;
        $\text{RegrMin}[i,g].\textsc{Observe}\left(\theta_{i,g}\right)$\tcp*{feed the local learners} \;
    }
}
\KwReturn{$\gamma$}
\end{algorithm}

\paragraph{Interpretation of Algorithm~\ref{algo:progressive-hiding}} Suppose that at stage $t$ of the algorithm, the policy  is $\mu^{t}_i$ for $i\in I_p$. Then for $i\in I_p$ and $g\in G_i$,
the local reward vector that is fed to the regret minimizer is $\mathbb{E}_{\delta_k,\mu_{-i}^{t},\mu_{-p}^t}[r_p(\textbf{h})|\tilde{\mathcal{X}}_i(\textbf{h})=g]$
with $k$ in $[\mathcal{A}(g)]$. We add a term proportional to the gradient of $-|| \gamma_i(g)-\mu^{t}_i(g) ||^2$ so that we optimize for the penalized criterion~\eqref{eq:criterion}.
It should be noted that thanks to the projection step, the algorithm output an implementable policy no matter the choice of $\lambda$ and $\tilde{\mathcal{X}}$. While the rational is different, we still note that the modification of a game payoff with a proximal term that makes the full criterion \textit{policy-dependent} is also observed in~\cite{perolat2021poincare}. 
In our case, however, the proximal term contains a projection, and is used to push the solution toward implementability. 

\subsection{Properties}

\begin{theorem}
\label{th:CFR}
Suppose $\tilde{\mathcal{X}}$ satisfies the perfect recall condition~\ref{eq:perfect-recall}.
     Then the overall regret is bounded according to the  relation
$
R^T
\leq \sum_{i\in I_p}\sum_{g\in G_i}R^{T,+}_{loc}(i,g)$.
\end{theorem}
Otherwise said, Theorem~\ref{th:CFR} states that CFR can be applied in the auxiliary game. However, note that the regret $R^T$ is defined with respect to the auxiliary game payoff.
Mirroring   Theorem~\ref{th:RIR} for Algorithm~\ref{algo:RIR}, Theorem~\ref{th:CFR}, with this bound on the regret in the auxiliary game,  quantifies what  Algorithm~\ref{algo:progressive-hiding} seeks. 
\paragraph{Proof Sketch for Theorem~\ref{th:CFR}}
The inductive proof borrows some aspects of~\cite{zinkevich2007regret} with the additional  penalty terms that need to be accounted for.

The next property  (which does not require perfect recall to hold)  allows sizing the distance between $\mu^t$ and  the implementable set. 

\begin{proposition}
    \label{corr1}
\begin{align*}
        \nicefrac{1}{T} \sum_{t=1}^T\left(\lambda^t\mathbb{E}_{\mu_p^t,\mu^t_{-p}}\sum_{i\in I_p}\bigg[||\mu^t_i(\textbf{h})-\text{Proj}^t(\mu^t_p)_i(\textbf{h})||^2 \bigg]\right)\leq R^T + 2 .||r||_{\infty}.
\end{align*}
\end{proposition}

While we leave the choice of $\lambda^{t}$ open, we report from our preliminary experiments that selecting a constant value with a grid search  seems a viable strategy.

An important assumption to guarantee low-regret with counterfactual regret minimization is that the player (or the group of player) satisfies perfect recall. 
In practice, to cope with games of large size,  the algorithm is often applied with an approximation of the game representation, which implies that perfect recall might not hold. 
Similarly, here, while we can provide guaranty when the relaxation induces perfect recall to hold, we also believe that for applications, an approximation of the perfect recall could also be envisioned for scalability concerns.

\section{Experiments}
\label{sec:experiments}

This paper focuses on games without perfect recall because we identify this setting as a relevant use case, since the method allows recovering Counterfactual Regret Minimization (CFR). In the competitive setting, a modified game can similarly be defined using either static information (hidden cards in the deck, dice results) or dynamic information (actions of the opponents) to improve the learning process, but this is left for future work.

This section reports on preliminary numerical experiments that complement the theoretical findings. We tested three implementations of Progressive Hiding on three different (team) games.

\subsection{Trade Comm}
\label{sec:tradecom}
We run two versions of progressive hiding on  \textbf{Trade Comm}~\cite{lanctot2019openspiel} because the game is elementary to solve for a human, but appears to be quite challenging for  learning algorithms~\cite{Sokota_Lockhart_Timbers_Davoodi_D’Orazio_Burch_Schmid_Bowling_Lanctot_2021}.  Furthermore, the game is parametrized by two parameters, which allows us to test different settings. 
The code is provided in the Supplementary Material. Of independent interest, the code closely aligns with the product game description, hence despite their equivalence in our context, the tree, and product representations lead to distinct coding paradigms. The no-regret learner we used for this game is FTRL with entropic regularization.

\begin{figure*}[htbp]
    \centering
    \includegraphics[width=1.\textwidth]{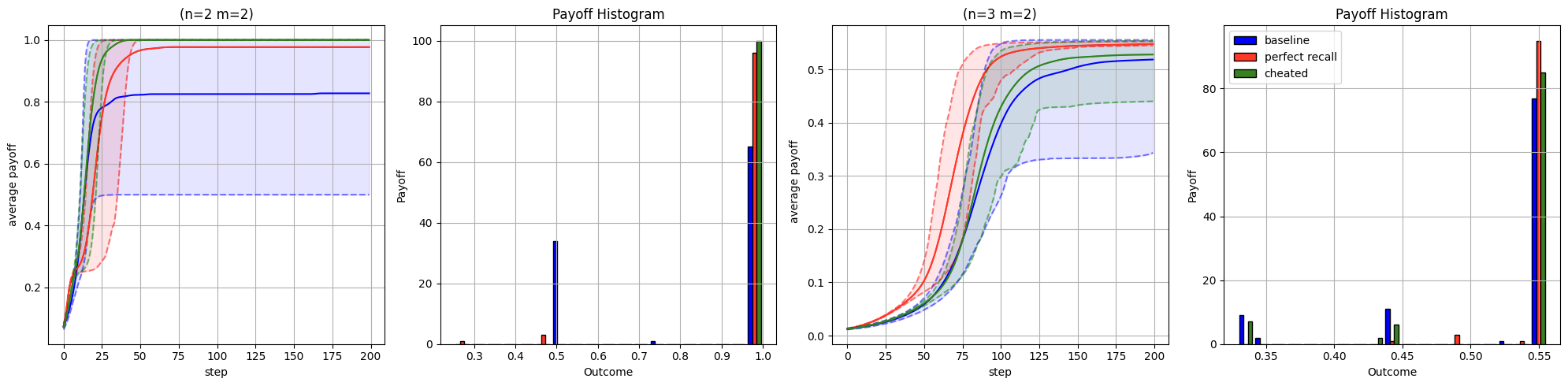}
    \caption{Learning outcomes ($\mathbb{E}_{\gamma^{t}}[r_t(\textbf{h})]$) distribution for the three information map \textsc{baseline}, \textsc{recall} and \textsc{cheated}   on \textbf{Trade Comm} with parameter  $(m,n)=(2,2)$ (left) and  $(m,n)=(3,2)$ (right).}
    \Description{Learning outcomes ($\mathbb{E}_{\gamma^{t}}[r_t(\textbf{h})]$) distribution for the three information map \textsc{baseline}, \textsc{recall} and \textsc{cheated}   on \textbf{Trade Comm} with parameter  $(m,n)=(2,2)$ (left) and  $(m,n)=(3,2)$ (right).}
    \label{fig:enter-label1}
\end{figure*}

Trade Comm is a common-payoff communication  game.
Each  of the two players randomly receives one of several items in $[n]$ ($s_1$ and $s_2$).
The first player communicates by choosing one out of several possible messages in $[m]$, denoted by $m_1$, which the second player observes.
Then, the second player also selects a message (in $[m]$), denoted by $m_2$, for the first player to see.
Both players then secretly request a trade involving one of the possible item combinations ($(d_1^1,d_2^1)$ and $(d_1^2,d_2^2)$).
A trade is successful if both players request to exchange their items for each other’s. They both earn a point if the trade works out, and none if it doesn’t.  Despite seeming straightforward, Trade Comm effectively highlights the challenges in same payoff games. 
We showcase that information relaxation offers a range of strategies through the choice of the information map $\tilde{\mathcal{X}}$ and the penalty schedule $\lambda^{t}$.

Formally, $P=1$,  $(s_1,s_2)\in \Omega=[n]\times [n]$,  $\mathbb{P}$ is the uniform distribution on $\Omega$, the product space $\mathbb{H}$  is 
\begin{align*}
\underbrace{([n]\times [n])}_{\Omega}\times \underbrace{[m]}_{\text{message 1}}\times \underbrace{[m]}_{\text{ message 2}}\times \underbrace{([n]\times [n])}_{\text{trade request 1}}\times \underbrace{([n]\times [n])}_{\text{trade request 2}}.
\end{align*}
The information map $\mathcal{X}$ is 
    $\mathcal{X}_1(h)=s_1$, 
    $\mathcal{X}_2(h)=s_2$, 
    $\mathcal{X}_3(h)=(s_1,m_1,m_2)$,
    $\mathcal{X}_4(h)=(s_2,m_2,m_1)$, 
the available actions map $\mathcal{A}$ is 
    $\mathcal{A}_1(h)=[m]$,   
    $\mathcal{A}_2(h)=[m]$, 
    $\mathcal{A}_3(h)=[n]\times[n]$, 
    $\mathcal{A}_4(h)=[n]\times[n]$, 
and, last,  the reward is $r_p(h) = [d^1_1 = s_1]\cdot [d^2_2 = s_1]\cdot [d^2_1 = s_2]\cdot [d^1_2 = s_2]$.
We can check that perfect recall is not satisfied for $n>1$.

We envision three possible information maps: the "original" information map $\mathcal{X}$, the "cheater" information map that reveals the private information, but does not record the message sent. This type of imperfect recall can be desired  for computational efficiency in the same spirit as game abstractions are used: 
 $\mathcal{X}_1^{cheat}(h)=(s_1,s_2)$, 
  $  \mathcal{X}_2^{cheat}(h)=(s_2,s_1)$, 
  $  \mathcal{X}_3^{cheat}(h)=(s_1,s_2,m_2)$,
    $\mathcal{X}_4^{cheat}(h)=(s_2,s_1,m_1)$, 
and a \textit{perfect recall} information map, that, as it names suggest, ensures perfect recall: 
   $ \mathcal{X}_1^{PR}(h)=(s_1)$, 
    $\mathcal{X}_2^{PR}(h)=(s_1,s_2,m_1)$, 
    $\mathcal{X}_3^{PR}(h)=(s_1,s_2,m_1,m_2)$,
   $ \mathcal{X}_4^{PR}(h)=(s_1,s_2,m_1,m_2,d_1^1,d_2^1)$. 
We insist that the information maps remain static during training. Progressive hiding is achieved entirely by adjusting the penalty parameter sequence. Both the cheater map and the perfect recall map could be used. The paper does not address which map is theoretically better, as this is beyond the scope of the current work.
However,  Theorem~\ref{th:CFR} suggests that  ensuring perfect recall is a reasonable approach.

The Appendix in the supplementary material contains more details on the experiments. 
% We provide instance of trajectories in Figure~\ref{fig:enter-label2}.
We compare the three information maps for $(n,m)=(2,2)$ and $(n,m)=(3,2)$ 
in Figure~\ref{fig:enter-label1}. The dashed lines correspond to the 10\% quantiles, and the full line to the average value, over 100 runs with randomized initial policy. 
We see that the methods  relying on progressive hiding beat the baseline on the two examples. Progressive hiding reaches the optimal payoff more often and faster than the baseline.  
It is notable that even \textit{cheated}, which consists in an information relaxation not big enough to satisfy perfect recall, still induces a clear improvement over the baseline. 
\subsection{Cooperative Matching Pennies}
\label{sec:matching_pennies}

\begin{figure}
    \centering
    \includegraphics[width=.15\textwidth]{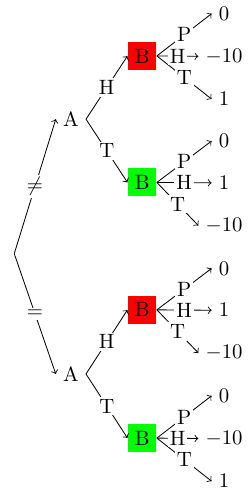}
    \caption{Tree representation of Cooperative Matching Pennies, introduced in Section~\ref{sec:matching_pennies}.
First a random state is sampled among SAME or DIFFERENT. Alice, the first player, observes the outcome of this random event and then chooses between TAIL and HEAD. Bob, the second player, knows Alice's choice but does not know the state of nature. Bob then makes his decision, choosing either TAIL, HEAD, or PASS. The payoff, indicated in the leaves,  is the same for both player. }
    \label{fig:matching_pennies_tree}
\end{figure}

The game is described in Figure~\ref{fig:matching_pennies_tree}.
We designed this game so that it is a difficult game for the agents, but the solution is easy to see. 
Indeed, Bob is incentivized to pass in the early learning stages, so that the learning stops. We  pitted a Monte-Carlo version of CFR versus its progressive hiding equivalent. We used a penalty parameter of 0.05 for progressive hiding. We considered the learning successful if, after 400 episodes, the algorithms reached an expected payoff greater than 0.95. We repeated the experiment 1000 times. While CFR never succeeded, progressive hiding succeeded $48\%$ of the time.

\subsection{Abstracted Tiny Bridge}

\begin{figure}
    \centering
    \includegraphics[width=.5\textwidth]{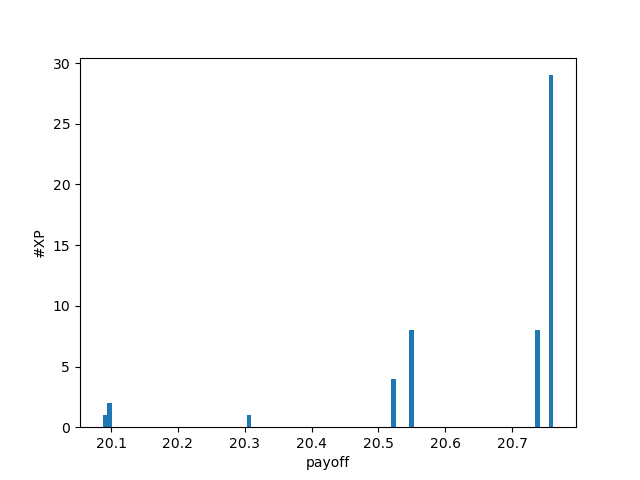}
    \caption{Distribution of the best maximal payoff obtained along the learning for each of the 50 training for Abstracted Tiny Bridge.
    According to \cite{Sokota_Lockhart_Timbers_Davoodi_D’Orazio_Burch_Schmid_Bowling_Lanctot_2021},  20.32 is the performance of the best joint policy that does not requires coordination.}
    \label{fig:bridge}
\end{figure}

This game is available in Open Spiel~\cite{lanctot2019openspiel}\footnote{see \url{https://github.com/google-deepmind/open_spiel/blob/master/open_spiel/games/tiny_bridge/tiny_bridge.h} for a brief explanation by the developers of the library}.
Abstracted Tiny Bridge is a simplified, cooperative version of contract bridge that preserves key strategic aspects. In this game, each player privately receives one of 12 possible hands. The players then engage in bidding to set the contract. The overall payoff is determined by the selected contract, the hand of the player who chose the contract, and the hand of the other player. The game's challenge lies in using bids to both communicate hand information and establish the contract, with fewer choices remaining as bidding progresses.
We tested progressive hiding with regret matching. 
Instead of setting  a value for the penalty parameter, we used a dynamic penalty parameter to control the expected payoff of the relaxed policy. 
The result of this experiment is summarized in Figure~\ref{fig:bridge}.
It is notable that we were able to achieve those results using only 75 episodes, while the authors in~\cite{Sokota_Lockhart_Timbers_Davoodi_D’Orazio_Burch_Schmid_Bowling_Lanctot_2021} used 10 million episodes for their baselines, and 100 thousands for their algorithm (CAPI).
We were unable to achieve the payoff of CAPI (close to 21, while our peak is at 20.76) though, still progressive hiding enjoys performance comparable to the other methods but with a microscopic budget of episodes.

\section{Discussion}
\label{sec:conclusion}
In this article, we show how information relaxation can be made compatible with learning in games. We build on no-regret learners to propose progressive hiding. 
We show that the algorithm is principled and showcase promising experimental results.

\paragraph{Limitation}
The main limitation of progressive hiding is its local nature, which stems from our definition of regret.
Another limitation of progressive hiding is its scalability, but we can be optimistic in the possibility to combine the method
 with Monte Carlo sampling~\cite{NIPS2009_00411460} and  deep learning approaches~\cite{pmlr-v97-brown19b,hennes2020neural}. 
 Last, a question that is not  addressed in this work is how to choose $\lambda^t$.

\paragraph{Further work}
A question that is not fully answered in this work is the choice of information map to perform the information relaxation. 
It seems intuitive that this question relates to the literature on game abstractions. 
Also,  while we introduced information maps (\textit{i.e.}, functions) to encode the player's knowledge for the sake of simplicity,
 it seems that information fields~\cite{carpentier2015stochastic} because they have a lattice and algebraic structure, would be a better mathematical tool if one wants to go further. Typically, an information field of interest is the smallest field greater than the original field that ensures that perfect recall is satisfied.
In this direction, it would be interesting to know if there exist informational properties other than perfect recall that allow for decomposition results.

\begin{acks}
BH would like to thank Michel De Lara  for introducing him to  Witsenhausen's model.
We thank the anonymous reviewers and the area chair for their constructive feedback.
\end{acks}

%%%%%%%%%%%%%%%%%%%%%%%%%%%%%%%%%%%%%%%%%%%%%%%%%%%%%%%%%%%%%%%%%%%%%%%%

%%% The next two lines define, first, the bibliography style to be 
%%% applied, and, second, the bibliography file to be used.

%\bibliographystyle{ACM-Reference-Format} 
\balance

%%%%%%%%%%%%%%%%%%%%%%%%%%%%%%%%%%%%%%%%%%%%%%%%%%%%%%%%%%%%%%%%%%%%%%%%

\end{document}